\documentclass{sigchi}

\newcommand\acksname{Acknowledgments}
\specialcomment{acks}{%
  \begingroup
  \section*{\acksname}
  \phantomsection\addcontentsline{toc}{section}{\acksname}
}{%
  \endgroup
}

\usepackage{balance}       
\usepackage{graphics}      
\usepackage[T1]{fontenc}   
\usepackage{txfonts}
\usepackage{mathptmx}
\usepackage[pdflang={en-US},pdftex]{hyperref}
\usepackage{color}
\usepackage{booktabs}
\usepackage{textcomp}
\usepackage{cite}

\toappear{\scriptsize Permission to make digital or hard copies of all or part of this work for personal or classroom use is granted without fee provided that copies are not made or distributed for profit or commercial advantage and that copies bear this notice and the full citation on the first page. Copyrights for components of this work owned by others than the author(s) must be honored. Abstracting with credit is permitted. To copy otherwise, or republish, to post on servers or to redistribute to lists, requires prior specific permission and/or a fee. Request permissions from permissions@acm.org. \\
{\emph{CHI '20, April 25--30, 2020, Honolulu, HI, USA.} } \\
\copyright~2020 Copyright is held by the owner/author(s). Publication rights licensed to ACM. \\
ACM ISBN 978-1-4503-6708-0/20/04\ ...\$15.00.\\
DOI: https://dx.doi.org/10.1145/3313831.3376647 }

\usepackage{microtype}        
\usepackage{ccicons}          

\usepackage{todonotes}

\def\plaintitle{Crafting, Communality, and Computing: Building on Existing Strengths To Support a Vulnerable Population}

\def\emptyauthor{}
\def\plainkeywords{asset-based; sensitive setting; HCI4D; global south; ICTD}

\makeatletter
\def\url@leostyle{%
  \@ifundefined{selectfont}{
    \def\UrlFont{\sf}
  }{
    \def\UrlFont{\small\bf\ttfamily}
  }}
\makeatother
\def\pprw{8.5in}
\def\pprh{11in}

\definecolor{linkColor}{RGB}{6,125,233}
\hypersetup{%
  pdftitle={\plaintitle},
  pdfauthor={\emptyauthor},
  pdfkeywords={\plainkeywords},
  pdfdisplaydoctitle=true, 
  bookmarksnumbered,
  pdfstartview={FitH},
  colorlinks,
  citecolor=black,
  filecolor=black,
  linkcolor=black,
  urlcolor=linkColor,
  breaklinks=true,
  hypertexnames=false
}

\begin{document}

\title{\plaintitle}

\numberofauthors{3}
\author{%
  \alignauthor{Aakash Gautam\\
    \affaddr{Virginia Tech, USA}\\
    \email{aakashg@vt.edu}}\\
  \alignauthor{Deborah Tatar\\
    \affaddr{Virginia Tech, USA}\\
    \email{dtatar@cs.vt.edu}}\\
  \alignauthor{Steve Harrison\\
    \affaddr{Virginia Tech, USA}\\
    \email{srh@cs.vt.edu}}\\
}

\maketitle

\begin{abstract}
In Nepal, sex-trafficking survivors and the organizations that support them have limited resources to assist the survivors in their on-going journey towards reintegration. We take an asset-based approach wherein we identify and build on the strengths possessed by such groups. In this work, we present reflections from introducing a voice-annotated web application to a group of survivors. The web application tapped into and built upon two elements of pre-existing strengths possessed by the survivors --- the social bond between them and knowledge of crafting as taught to them by the organization.  Our findings provide insight into the array of factors influencing how the survivors act in relation to one another as they created novel use practices and adapted the technology. Experience with the application seemed to open knowledge of computing as a potential source of strength. Finally, we articulate three design desiderata that could help promote communal spaces: make activity perceptible to the group, create appropriable steps, and build in fun choices.
\end{abstract}



\printccsdesc

\section{Introduction}

Human trafficking is an acute problem prevalent across the globe. Prevention has received considerable attention;
however, support for the well-being of trafficking survivors is limited and research on their long-term prospects and well-being is even more limited  \cite{dell2017helping, oram2016human}. 
In Nepal's context, an estimated 15,000 women are trafficked within and outside Nepal annually \cite{ktmNews2019}. 
Non-governmental organizations (NGOs) are the major actors involved in repatriation, rehabilitation, and reintegration of sex-trafficking survivors \cite{kaufman2011research, laurie2015post}.

Our work is situated in one such anti-trafficking NGO in Nepal and explores prospects for sex-trafficking survivors living in a protected home. 
As part of our work, we have been exploring ideas of designing socio-technical systems that may help the survivors achieve ``dignified reintegration'', a phrase commonly used by the NGO.
However, reintegration into society is challenging for a myriad of reasons including that many in Nepali society believe that survivors bring disgrace not just to their families but to the entire community \cite{mahendra2001community, simkhada2008life, poudel2009dealing, joshi2001cheli, sharma2015sex, richardson2009sexual}.
The survivors are vulnerable in a number of ways: most are young, uneducated, impoverished and have experienced traumatic ordeals, many have already been shunned by their families, others might be shunned in the future, and those we worked with are dependent on the NGO that has rescued them. 
Further, NGOs are subject to complex political, economic and cultural forces that drive their programs. 
Similar programs and interventions have been found to be helpful but concerns remain about their effectiveness \cite{crawford2008sex, kaufman2011research, poudel2009dealing, sharma2015sex}.

In our earlier ethnographic study in the same context, we had uncovered two strengths possessed by the survivors: crafting skills taught by the NGO and a social bond among the survivors \cite{redacted}.
We argue the need to build upon these elements of strengths to move towards long-term, dignified reintegration. 
In this paper, we present findings from a step that we took in that direction by introducing a voice-annotated web application called Hamrokala to a group of survivors. 

Hamrokala was contextualized around crafting and was designed to promote interactions between the survivors.
It allowed survivors\footnote{The survivors addressed each other as ``sisters''. The researcher addressed the survivors as ``sisters'' as well. To match this nomenclature, we shall henceforth call the group we worked with ``sister-survivors''.} to post crafting items as if for sale on the internet.  
The setting was made communal by the sister-survivors' behavior but the web application was also tailored to build on this, allowing communal activities such as drawing and sharing design ideas, seeing other sister-survivors' crafts, and commenting on them. 

Further, the design of any technology for use in this context is not straightforward. 
A typical technology interface assumes some level of familiarity with text but an estimated 85.9\% of the survivors in Nepal have \emph{never} been to school \cite{nhrc2018}.
Similarly, survivors have limited digital fluency.  
In fact, the NGO's previous attempt to introduce technology to the survivors had failed, with the survivors reporting being overwhelmed \cite{redacted}.  
Care was placed on making the application approachable by following design principles for low-literacy populations \cite{ghosh2003design, huenerfauth2008developing, medhi2011designing, medhi2006text}.  
Additionally, voice annotations were prioritized. 
The annotations, played in a female voice in Nepali, were longer and more elaborate than usual, and used phrases tailored to the population. 
The voice annotation thus served to relieve the pressure to read. 

This study reports on a ten-day-long training session with nine sister-survivors. 
Our contributions are three-fold.
First, we see Hamrokala and the workshop as a potential contribution.
We argue that presenting a tailored computing application contextualized around a familiar activity facilitated adaptation and appropriation of technology. 
Second, a result of our study is that we now believe that, with sufficient care, computing could be a resource, working with the other two existing strengths, to support long-term reintegration. 
We add to existing scholarship by extending considerations beyond acceptance of the technology to its operation in building upon and increasing existing strengths. 
Third, we identify three qualities in spaces to promote communality for such vulnerable populations: (1) make activity visible to the group, (2) make every step appropriable by someone in the group, and (3) build fun choices in the socio-technical system even when a focus on economic well-being might indicate a more restrictive approach.

\subsection{Crafting as a Means to Support Survivors in Nepal}

While some trafficking survivors choose to avoid NGOs after repatriation, many depend on them for support.  
The survivors who receive support from NGOs typically undergo what they call ``rehabilitation'' and ``reintegration''. 
Rehabilitation is a contested term because it may be understood as signifying that the survivors themselves are somehow responsible for being trafficked and now have to be changed before rejoining society \cite{laurie2015post}. 
However, it captures the real problems that survivors face not only because of their interrupted lives but also because of the stigma that Nepali society places upon them.  
NGOs provide survivors with some form of psycho-social counseling, medical checkups, and legal assistance \cite{kaufman2011research}. 
NGOs may also try to reintegrate the survivors with their families. 
Additionally, whether reintegration into their former lives is possible or not, 
the NGOs try to strengthen the survivors' economic self-sufficiency through skills-based craft training, typically making local handicraft such as \emph{Pote}, a Nepali glass-bead necklace, and knitted bags, footwear, and scarfs.

Research has highlighted systemic challenges faced by trafficking survivors such as in gaining citizenship --- which is necessary to have a bank account or own property, and is largely inaccessible to those disowned by their families \cite{richardson2016women} --- or in gaining sustainable livelihood \cite{crawford2008sex}.
The prospects through crafting have been found to be limited. 
For example, Crawford and Kaufman \cite{crawford2008sex}, in following up with 20 survivors found that while 17 had returned to their villages after leaving the shelter homes, 11 were involved in some form of income-generating work out of which only two (2) ran tailoring shops, work loosely related to the crafting skills provided by the NGOs.

Despite the limitations, there are pragmatic reasons for NGOs to support crafting including that it can be taught even when resources are extremely limited, it is culturally-acceptable, and it is valued, understood and attainable by all the survivors.  
Unlike many other skills, crafting does not have to be taught by cohort, which is a critical factor given that there is no fixed time when survivors join or leave the shelter homes. 
The NGO in this study invested in trainers and resources to operate the handicraft space because they thought crafting would be therapeutic and indeed the sister-survivors have reported it as helpful \cite{redacted}.
Selling crafting products also brings in some money for the NGO and the individual survivors.

\subsection{Reflexivity and Commitments}

The first author is a male who was born and raised in Nepal. 
As a male conducting research in a largely female setting, he stands out. 
As Nepali, he brings awareness of cultural norms and practices that make interactions amenable in this context, he, as a researcher with great privilege, is an outsider. 

All three authors (2 male, 1 female) are based in an academic institution in the United States. 
A part of our research has focused on issues of power and authority.
We find the patriarchal values and norms that are prevalent in Nepal problematic, and through this work, we see ourselves trying to support a group that has been marginalized and made vulnerable by exploitation. 
We are aware that reintegration is not our struggle. 
We want to ``stand with'' \cite{tallbear2014standing} the sister-survivors.


An additional factor is our relationship with the NGO. 
In the past three years, we have been building a relationship, going ``beyond ethnography'' \cite{brereton2014beyond} with gestures such as helping to maintain their website.  
It is important to treat the NGO and its decisions respectfully, not only because we rely on them for access to the sister-survivors, but also because of limitations on our knowledge. 
Our working assumption is that the NGO is operating in the best way it sees possible with the resources it has. 
While continuing with crafting can be seen as creating moral quandaries, especially if the survivors do not enjoy crafting activities, crafting can also be seen as a crucial element of existing practice in this setting.

\section{Related Work}
\subsection{Building on Existing Strength}  

We come into this study with an asset-based approach \cite{kretzmann1993building, mathie2005driving} that aims to build on the sister-survivors' existing strengths.  
This is in contrast to need-based, deficit-focused approaches. 
Kretzmann and McKnight \cite{kretzmann1993building} argue that a deficit-focused approach can lead to several problems including when the community internalizes dependencies and loses the agency to bring about change.  
This, in turn, can incentivize potential change-makers to maintain the dependency by misrepresenting problems so as to attract resources.  
In fact, research work highlights how anti-trafficking NGOs, including those in Nepal, often take a deficit-focused approach leading them to be termed as being part of ``the rescue industry'' \cite{agustin2007sex, laurie2015post}.

Asset-based approaches have begun to gain interest within our research community (e.g. \cite{cho2019comadre, ismail2018engaging, karusala2017care}). 
A key task in this approach is to identify assets possessed by a community, and seek ways to build on them, as we have begun to do.
We had previously identified two strengths that the sister-survivors possessed: (1) crafting skills that could be a source of livelihood and (2) social ties with one another \cite{redacted}.

However, both these strengths were fragile.  
On the one hand, the sister-survivors wanted to show and sell their crafts to others and persevered through difficulty to learn to craft. 
On the other hand, they were also concerned about dwindling sales, limited access to the market and the limited long-term feasibility of crafting. 
Some also found crafting boring at times. 
Similarly, the sister-survivors expressed concern regarding the lack of opportunity to engage with other members of the community and the potential loss of connection with others once they left the shelter home.

By introducing Hamrokala, we leverage technology to build upon and increase the sister-survivors' existing strengths. In particular, we add elements of sociality and fun to their crafting practices, provide an avenue to situate their crafting practices within the wider economic realities, and potentially widen their access to market to sell handicrafts.

\subsection{Social Bonds and Care}

Sister-survivors displayed social bonds with one another. Their observed behavior and comments can be interpreted from a number of overlapping but different frames: as expected behaviors in collective societies, as reflecting personal attachment, or as an acknowledgment of community or communality. They are also living in a setting that may invoke the notion of care. A central aspect of our study context involves the NGO caring for the sister-survivors by having rescued them, providing a supported living situation, and offering training in crafts.  

Care, which at its most general ``is inevitably to create relation'' \cite[pp. 198]{de2012nothing}, has been explored within our research community, in various contexts such as disaster relief \cite{wong2017social}, data science practices \cite{zegura2018care}, and the experiences of a Wikipedia contributor \cite{howard2019ways}.
Toombs \textit{et al.} \cite{toombs2015proper}, in their ethnographic inquiry about a makerspace show how despite tensions with the underlying neoliberal, individualist ethos, care was enacted both implicitly and explicitly. They argue that care is essential for sustenance of the makerspace, by building ties with one another and the larger community \cite{toombs2015proper}. 
Similarly, on studying an underserved after-school learning center in India, Karusala and Vishwanath \textit{et al.} \cite{karusala2017care}, present care as a resource that can help foster interdependency, community, and a sense of ownership. Taking an asset-based approach, they present ways in which technology can be used to build on the existing caring practices to extend care in the learning center. 

Mol \textit{et al.} caution that ``care is not an innate human capacity'' and add that care ``may be adapted and improved along the way when they are attended to and \emph{when there is a room for experimentation}'' \cite[pp. 14]{mol2010care} (emphasis added). 
Indeed, care can be entangled, conflicted and ``non-innocent'' \cite{murphy2015unsettling}.
Attempts to care can push the carer's agenda, which may diminish the values of those who are being cared for \cite{murphy2015unsettling}. 
Thus a caring environment should involve an ongoing negotiation of values \cite{karusala2017care, labonte2001capacity, mol2010care, putnam1994productive}.

In the current case, neither we nor the sister-survivors have significant power to negotiate values with the NGO, which acts as the central actor bringing together the sister-survivors with one another and offering us access to them. 
The central role of the NGO means that we prefer to describe the sister-survivors' social bonds with one another as ``communality'', to avoid relying on the enduring existence of strengths which may be highly contingent on particulars of the circumstances.  
We hope that the social bonds can sustain an interdependent world for the survivors, but we cannot know this.  
We have only observed enough to attempt to increase the ways that such bonds can be negotiated and perhaps strengthened. 
One hope is to enhance the room for experimentation in their lives.

\subsection{Communal Use of Technology in the Global South}

HCI4D and ICTD literature highlights various socio-economic barriers to technology use and adaptation (e.g. \cite{chigona2008using, medhi2007optimal, pritchard2013digital,  sambasivan2010intermediated}).
Women are additionally limited by a range of barriers such as restricted physical mobility \cite{mudliar2018public} and privacy and security concerns \cite{sambasivan2019they}. 
It is worth reiterating here that technology alone cannot overcome the social, economic and cultural barriers faced by women and marginalized groups in the Global South.

Communality has been explored as a force in overcoming some of these barriers \cite{ahmed2015suhrid, ratan2009kelsa, vashistha2019threats}.  
For example, Kumar and Anderson \cite{kumar2015mobile} highlight the roles that children play in teaching their mothers and aunts to use technology. 
A few others have emphasized the value of playful, ludic experiences for entertainment consumption \cite{smyth2010there} or sharing information through voice-manipulation \cite{raza2012viral}.
Similarly, Johri and Pal \cite{johri2012capable} constitute a small but growing voice arguing the need for ``capable and convivial design'' in HCI4D/ICTD, which includes supporting people ``to interact and form relationships with other people'', emphasizing the importance of relationship building.

Given that sister-survivors leaving the shelter home are likely to have minimal or no privacy, promoting communality through the provision of cell phones could put their lives and well-being at risk. 
With that avenue effectively shut down, we turned to other ways to build on their strengths.
We observed sister-survivors providing help to one another when working on crafting, and engaging in the shared venting of frustration. Additionally, formulating discussion tasks as communal appeared to allow considerable participation and appropriation \cite{redacted}. 
One design proposition was that a sufficiently communal orientation to the web application and workshop might succeed where a more standard approach had disappointed.

\section{Methodology}
The design of Hamrokala was based on the findings from our earlier ethnographic study and was considered as an option to build upon the sister-survivors' strengths.
When the NGO had introduced computers and Photoshop, the sister-survivors had rejected the system, expressed feelings of being overwhelmed, and the NGO came to believe that computing was not attainable for the sister-survivors \cite{redacted}. 
This led us to design with the incremental step of exploring an application relevant to a familiar activity. 
In this study, we examine the interaction that ensued upon the introduction of Hamrokala. 
Hamrokala and the workshop may be thought of as a design probe \cite{fitton2004probing, hutchinson2003technology} in that we both observed interaction with it and solicited participant responses, evaluations, and suggestions throughout.

\subsection{Study Context}
\subsubsection{The Partner Organization}
The anti-trafficking organization we partner with was founded by a group of sex-trafficking survivors more than 15 years ago. 
It employs around 100 staff members, many of whom are themselves trafficking survivors. 
We call this organization a Survivor Organization (SO). 

SO conducts three major kinds of programs, including the focus of this study, the rehabilitation and reintegration program, which offers protected-living homes (shelter homes), skills-based training in handicrafts, and reintegration through the provision of jobs and/or reunification with families. 

SO allows sister-survivors to stay in the shelter homes ``as long as they need'' which seems more responsive to the individual needs than the fixed duration (typically 6 months) programs in some other NGOs \cite{gautam2018participatory}. 
Survivors have been reported to have left SO's shelter homes within 2 months; SO's programs, such as providing training in handicrafts, and our own interventions have to be cognizant of this flux.

\subsubsection{Participants}
Ten sister-survivors were being trained in the handicraft workshop that was housed within SO's main office as part of the skill-based training program. 
Nine of the ten sister-survivors, between 13 and 23 years old, participated in our study. Eight of them lived in the SO's shelter homes. One (S6) had started living outside of the shelter home. 

On average, this group had more years of formal schooling than the group we had encountered in our earlier study \cite{redacted}.  
Four of the sister-survivors were voluntarily attending a ``morning'' school from 6:30 a.m. to 9:30 a.m. Four others expressed plans to rejoin schools. 

Only two of the nine sister-survivors had ever owned a mobile phone. 
No one in the shelter home was allowed to own phones and S6 had a simple phone. 
Three sister-survivors (S1, S5, and S6) had used a computer at least once. 
None in the shelter homes had access to a computer or the Internet.

\subsubsection{Hamrokala: The Web Application}
\label{hamrokala}

Hamrokala (``Our Craft'') was contextualized around crafting.
An earlier group of sister-survivors had expressed a desire to showcase and sell their crafts but were worried about its limited demand in the local market \cite{redacted}.
So we built features that allowed them to express their thoughts about the handicrafts, post those for sale, and share it with other sister-survivors.
They could also check the inventory, watch videos of experts creating similar handicrafts, and draw and share sketches.

An audio file played whenever a user hovered over a navigational or informational element in the web application. 
The audio files contained narration spoken by a native Nepali woman.
She used long descriptive words and colloquial Nepali phrases as if in conversation rather than typical computer-based labels. 
For example, instead of using unfamiliar words like ``login'' and ``logout'' which are typically used in Nepali websites, we used ``to go inside'' and ``to go outside'' in the written form.  
Further, the spoken version was even more naturalistic, saying, ``If you want to go inside, press here''. 
Voice was conceived as a communication rather than an efficient information source or a reading lesson. 
The written form was often shorter than the voice form.

This approach was underscored socially. During the sessions, we explained that login is ``similar to how you come inside the workshop to work on crafts'', and logout is ``similar to how you leave the workshop after your work is done''.

\begin{figure}
\centering
    \includegraphics[height=3.1cm, width=0.42\textwidth]{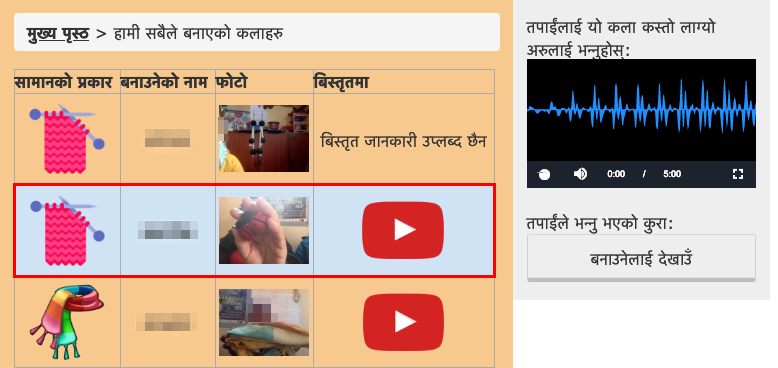}
    \caption{Users could share their crafts which would be visible to all members of the community. Members of the community could click on a shared item to see the details and leave an audio comment.}
    \label{fig:communalInterface}
\vspace{-3mm}
\end{figure}

Hamrokala sought ways to build upon and promote interactions and communality between the sister-survivors.
The sister-survivors could share items that they had posted for sale which were listed on a single page for everyone to see. 
The system supported submitting audio comments on those shared objects (Figure \ref{fig:communalInterface}).
We implemented a similar feature around drawing that allowed users to draw, import, edit, and re-share drawings.

Hamrokala was hosted on a local server accessible through a password-protected ad-hoc network.
This configuration helped us convey, in general terms, to the sister-survivors that all their data remained within the room and was protected.

\subsubsection{Workshop Sessions}

After receiving institutional review board (IRB) approval, we conducted ten two-hour sessions on computers and Hamrokala during January 2019. 
The first two days were used to talk and emphasize sociality, and to ensure a high level of comfort and agency. 
The first author introduced himself, our project, history with the organization, and read out the IRB consent document. 
At the beginning of the sessions, he reminded the sister-survivors of their right to drop participation at any time without incurring any penalty and asked for their permission to record audio, and started the recording only after he received permission from all.
Computers were introduced on the third day. Sister-survivors shared four laptops in groups (three groups of twos and a group of three participants).  
On the final day (tenth day), the sister-survivors were paired with a staff member to whom they explained the web application. 

The sessions were conducted in an environment that was familiar to the sister-survivors, in a room next to SO's handicraft workshop. 
Throughout the sessions, the sister-survivors left the room, came back, and moved around, which we felt showed that they knew they were free to come and go. 
The staff members were not present during the sessions, except on the final day.
The laughter, participation, and freedom of movement that the first author observed during the sessions suggest that the program succeeded, to some extent, in staying within the sister-survivors' comfort zone.

\subsection{Data Collection and Analysis Method}

Field notes and audio were recorded during the workshop sessions. 
We also recorded a video of the sister-survivors' screens when they explained the web application to a staff member on the final day of the session.
The audio and video were translated into English and transcribed by the first author.

The transcripts were first analyzed using an inductive process because of the unique setting of the study.  
In the first cycle of coding, the first author performed ``eclectic coding'' \cite{saldana2015coding} in which he conducted both descriptive and process coding.  
Descriptive coding closely follows and summarizes the text.  
Process coding searches for ``ongoing action/interaction/emotion taken in response to situations'' \cite[pp. 96]{saldana2015coding}.

The first two authors met regularly and discussed the codes and the transcript text, as suggested when doing ``solo coding'' \cite{saldana2015coding}.
Examples from the first cycle of coding are ``expressing delight in being able to share videos'', ``refusing to critique crafts'', and ``asking for help''.

In subsequent rounds of meetings all three authors discussed and combined the codes, leading to the emergence of higher-level codes such as ``awareness of the work'', ``elaborating the system's action'', and ``declaring a norm''. Subsequent discussion distilled these to 21 higher-level codes.  
Although the analysis of data started from an open position, we had a goal of building upon existing strengths.  Therefore, at this point, we began to introduce questions into the process about how the codes revealed strengths that could provide a sufficient and reliable basis for further work.  In the subsequent two rounds of coding, higher-level codes emerged around three key resources: computing, communality, and crafting. The findings section is centered around these three resources. 

Trust is central; we need to earn the trust of our study participants and be true to that. The sister-survivors expressed pride in having learned to draw on the computer but some expressed hesitation to share their drawings to people outside the group. So, while we report the number of drawings, we have not analyzed nor shared their drawings.

\section{Findings}

We came into the study with the goals of understanding whether Hamrokala together with the related workshop could work with this population and seeing, if, and how, it could be leveraged to build upon the sister-survivors' existing strengths. 
Our findings revolve around the three elements of strengths: computing, communality, and crafting.  

\subsection{Computing}

Computing has a dual status in this project as the putative mechanism of change and as a potential source of strength.  Evidence of use of the system in general and of the particular mechanisms designed to promote use are particularly important in evaluating whether computing had a chance of succeeding as a mechanism of change.  

The sister-survivors created a large volume of digital artifacts, including 47 clips of videos about their crafts of which 32 were included as part of an artifact for sale. 
They further had eight artifacts descriptions saved as drafts with five more item-description videos. 
They shared 24 instances of artifacts for comment by others in the group. 
They also made 45 audio comments and 38 drawings.

\subsubsection{Initial Orientation Towards Computers}

All nine sister-survivors expressed a positive attitude towards technology. They had a desire to engage with family and friends using technology. 
For example, S7 envisioned a future where technology was seen playing a role connecting her with her family and friends, ``\textit{I want to use computers to meet friends through the Internet and also talk to people in my family.}''
The role of technology in their envisioned future was not limited to personal connections but also towards achieving professional aspirations.
S2 wanted to run a dance school and she saw computing as a resource where she could ``\textit{watch songs and videos and, from that, learn }[dance] \textit{steps.}'' 
Similarly, S6 wanted to learn photo and video editing that she thought would be beneficial when she becomes a trekking guide.

\subsubsection{Challenges Using Computers}

This positivity was a component of their success using the system, yet there is no reason to doubt that other sister-survivors in the past also started out positively.  Equally critical is that the sister-survivors had limited experiences of actually using computers, a situation that can lead to rapid disillusion and self-deprecation \cite{vines2017our}.

The sister-survivors expressed initial difficulty using the keyboard and touchpad, factors that could have led to the earlier rejection. 
However, the attention paid to making the interface approachable and building on the existing communal orientation of the sister-survivors appeared to have paid off; unlike in the previous experience, by the fourth day, all sister-survivors were able to use the touchpad comfortably enough to draw sketches like flowers and hills.   

Their discomfort was mitigated but not entirely eliminated. 
S2, on the final day, expressed both the achievement and the difficulty, ``\textit{With the sisters helping, it }[computing] \textit{was a fun, easy thing. Now }[showing it to the warden] \textit{I got scared and my hands shivered.}'' 
S2 may have been scared but the warden was impressed exclaiming, ``they learned all this in two weeks!''

\subsubsection{Using and Improving Voice Annotation}

Using detailed language, the voice annotation explained functions that could be performed within the link such as ``to add a new item, press here''. 
In initial use, the sister-survivors relied upon this aspect of the voice annotation to identify where they wanted to go, and also support others in navigating.
For example, on the third day, we heard S1 helping S2 who found it hard to read Nepali text by suggesting her to ``\textit{go there where it says, `to draw press here'.}''

The elaborated voice annotations seemed to help clarify the actions of various elements of the application. On the final day, all the nine sister-survivors used metaphor to explain login and logout as heard in S7's explanation, ``\textit{We put in the code number }[password]\textit{ here. Then we press this to go inside.}'' 

Other elements, especially where we had used literal Nepali translations, were harder to comprehend and required adjustment. For example, we originally used \textit{tippani}, the Nepali translation of ``comment'', to signify an action that allowed users to leave audio comments about a shared craft object.  The sister-survivors did not know what \textit{tippani} signified, variously speculating that it probably meant, ``\textit{all the materials that are required to make it}'', ``\textit{the estimated price}'', and ``\textit{what is good about it}''. 
This led S9 to exclaim, in frustration, \textit{``I have not even heard these words } [before]\textit{ so don't know what to do.''}  These expressions of frustration were turned into an opportunity for mutual learning when the researcher pointed out that the word was physically next to particular craft items on the screen (see Figure \ref{fig:communalInterface}) and asked ``What else could be said about a craft?'' 
This led the sister-survivors to connect \textit{tippani} to ``saying what is nice or not nice''. 
The researcher then changed \textit{tippani} to this phrase in the written and spoken interface, and noted that, much later, the sister-survivors used this phrase to explain the web application to the staff member.

\subsubsection{Evolving Attitudes Towards Computing}

Novelty effects with technology use have been well documented (e.g. \cite{consolvo2008activity, molapo2017video, johnson2016reflections}). Consistent with this, the sister-survivors started out positive and their enthusiasm continued despite challenges.  We frequently found ourselves having difficulty in ending the sessions within the stipulated time to the chagrin of the staff members who had to wait a long time to accompany the sister-survivors to the shelter home.

However, there were indications in their behavior and expression that their interest went beyond novelty towards the perception that computing could constitute a real asset.  
What they were enthusiastic about changed over time. In the beginning, the sister-survivors were thrilled when the voice annotation played in Nepali as they moused over the HTML elements. 
They repeatedly played the audio files without taking any action during the initial introduction, often mimicking the voice, including the intonation. 
They seemed to be engaged in active learning about the relationship between the screen and the sounds.  
As the sessions progressed, we found that sister-survivors seem to have stopped relying on the voice annotations.
They went back to the voice when they had difficulty with the interface but even then they played the audio file just to the point that they remembered the page section.

At the same time, the sister-survivors also expressed a sense of pride in being able to learn to do things on the computer.
S8 appreciated the fact that she learned to draw, ``\textit{I liked it }[computer]\textit{. I didn't know how to draw earlier but now }[I] \textit{can draw a little bit.}''
In addition, S2, S6, S7, and S9 mentioned that they learned how to speak about a craft as S2 explained, ``\textit{I learned a bit about how to say it and what all to say about a craft. And also how we could possibly run a business.}'' 
They also expressed interest in learning through feedback from people outside of the community. 
S3, for example, expressed, ``\textit{I want to show whatever skills I have, how so it may be because, for any of my bad drawings, I would get feedback and move forward on improving it}''. 
This sentiment of learning was appreciated by S6 who suggested to S8 to ``\textit{listen to that}''. 
S3 later reported, ``\textit{I felt increasingly that I could} [draw]''.

\subsection{Communality}

We hypothesized that the existing social bonds between the sister-survivors could be used to aid in the take-up of computing as well as be strengthened by the activity of computing.  

\subsubsection{Sharing Work}

Pleasure in communality was manifest in the sister-survivors' orientation towards Hamrokala's sharing features.  
``Did it come?'' was a commonly used phrase to confirm that the artifact such as drawing or comments could be seen by others. 
S9 was elated in being able to share and see videos of crafts:
\begin{quote}
    S9: \textit{I am very happy. I felt we know how to do it and that's wonderful} [laughs] \textit{I was also happy to see others' work. This was fun, even if we are all here, we don't have to go over to their side} [of the table] \textit{to see.}
\end{quote}

Similar pleasure was expressed regarding sharing drawings. 
S7 remarked, ``\textit{I did as much as I knew about drawing. If I sent it here, other people saw whatever I had done and I could see what others have sent, and to be able to see those was extremely nice.}'' 
The sister-survivors expressed a desire to both ``send'' their work to others as well as to see others' work. 

While some of this is presumably due to the novelty of computation, the sharing --- whether by the communication features of Hamrokala or just by what was seen as a collective endeavor --- was an explicit source of pleasure.  Such pleasure can be important in the success of the intervention \cite{smyth2010there}.

\subsubsection{Collective Behaviors}

Pleasure was also arguably present in other collective behaviors, which were more similar to what we had previously seen in their collective crafting practices \cite{redacted}. 
The sister-survivors moved around to help each other or ask for help. 
Sometimes they reached out to individuals, while at other times they asked the entire group. Support-seeking ranged from seeking information about the craft to help in using the web application. After S7 moved over to S9's side of the table, she explicitly asked for help in understanding a page in the web application:
\begin{quote}
    S7: \textit{Please teach me how to do this} [comment] \textit{from the start. I haven't understood anything about this.}
    
    S9: \textit{This is only for commenting.}
    
    S7: \textit{How did you comment? I have understood till here but I did not understand this part.}
    
    S9: \textit{Here you can see all the crafts that others have made and shared.}
    
    S7: \textit{Yes}
    
    S9: \textit{Here you have to press} [clicks on a shared craft] \textit{who you want to comment ...}
\end{quote}

Other times, the sister-survivors jumped in to help without being explicitly asked. 
For example, S7 seeing S8 in distress while drawing asked,  ``\textit{What happened?}'' to which S8 pointed to the screen where a Bootstrap Modal, a popup dialog box, had appeared. 
S7 then suggested, ``\textit{Do an into} [press the cross mark]''. 
S9 joined and helped S8 to draw, ``\textit{Press this} [the touchpad button] \textit{with one hand and you can draw with the other} [hand]'' and further provided encouragement as S8 followed her advice, ``\textit{You are making it very well}.''

Communal behaviors were not confined to collective problem solving but also involved encouragement in creating new practices. For example, S8 mentioned that whenever she tried to speak she could not help herself from laughing so she thought singing may be better for her.
Upon S9's encouragement, S8 sang a song as part of the item-description video. 
Similarly, S8 had earlier asked for scaffolds in the form of questions that she could write the answer before speaking.
The researcher explored a question-and-answer model where he asked her questions about the craft and she answered. 
This, she reported, helped her. 
He later observed fellow sister-survivors (usually S9) asking questions which S8 answered as she made her item-description videos.
Similarly, S6 too wanted scaffolds and was willing to support others. 
In fact, out of the 47 item-description videos, S6 can be seen or heard helping fellow sister-survivors in seven of them suggesting the importance of leadership.

\subsubsection{Individual Boundaries and Ownership}
Sister-survivors differed in where they were comfortable manifesting individual ownership.  
When discussing the possibility of sharing their anonymous drawings with others outside of the group, S6 reasoned, ``\textit{No, because} I \textit{would know that I have kept a drawing and know that I have done that drawing and because of that} [I would not want to share].''  
At the same time, S9, for example, showed little reluctance.  She wanted to share her work and created nine item-description videos that manifested personal ownership by showing herself holding the artifacts that she had made. 

\begin{figure}
\centering
\includegraphics[height=1.8in]{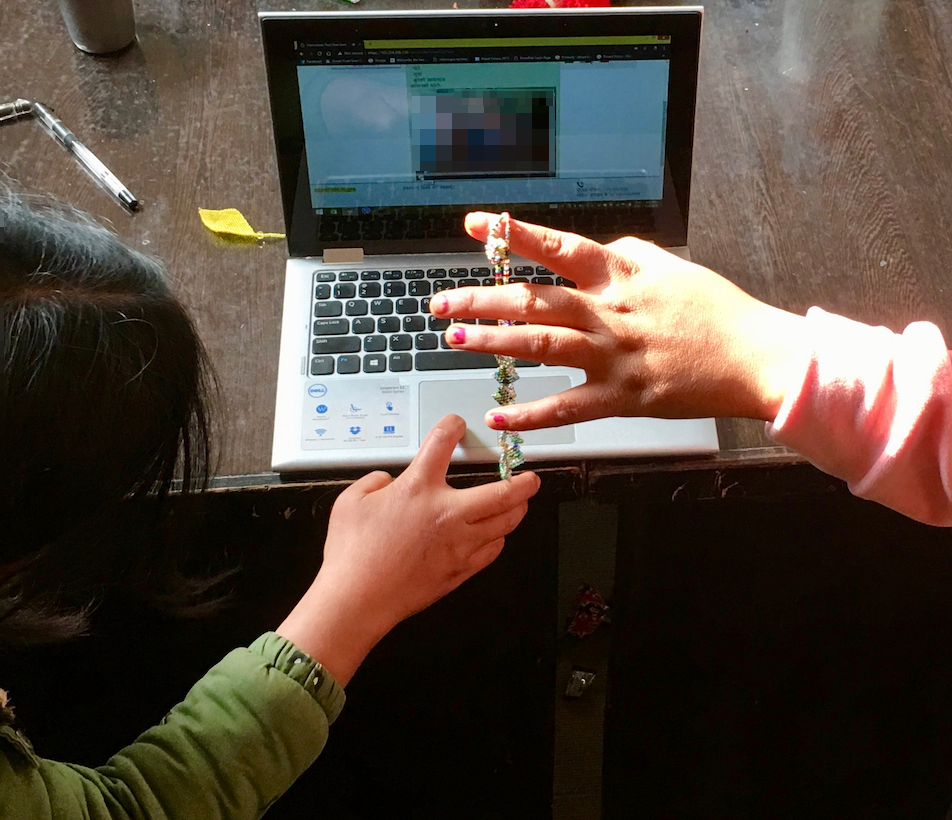}
\caption{S5 helping S4 to record an item detail. Most of the sister-survivors hid when capturing videos because of the fear of being identified as a trafficked person. Blurring added to protect their identity. }
\label{fig:uploading}
\vspace{-0.4cm}
\end{figure} 

However, fear of stigmatization meant that sister-survivors were not able to make free choices in this arena. S1, S5, S6, and S7 were hesitant in opening the camera to record video and thus they used the webcam cover during recording. Even S9 suggested that the webcam cover would help share their work with others outside the community:
\begin{quote}
    \textit{With only our voice, without the camera, it will be harder to recognize us. With only voice, it becomes unclear who spoke. When we are the only ones, when we know who all are, we can know whose voice it is. But when} [the video] \textit{goes out and speaks, it won't be easy to recognize who is the one speaking} [in the video].
\end{quote} 

There were boundaries within the group as well. 
All the sisters-survivors expressed respect for others' drawings. S1 refused to draw on top of S9's drawing because it ``would be ruined''.
Concern for boundaries could also be heard in S2's mixed feelings about the drawing feature, ``\textit{I liked it }[the drawing page] \textit{but I also didn't like it. I didn't like it because I could take others' nice drawings and ruin it. I felt I was ruining it. It was nice in that our drawing was going and I could see others' drawings.}'' 
This led the group to explicitly establish a rule: to not draw on top of someone else's drawing. 
S6, on the final day, expressed this sentiment, ``\textit{It is especially important that you do not draw on top of it} [someone else's drawing]. \textit{This} [the feature] \textit{is just here to show that it can be done.}'' 

When we mentioned that they would be drawing on a \textit{copy} of the original drawing and that the original drawing would remain intact, S6 clarified that ``ruining'' was not about physical damage but rather an artistic one, that of showing respect for the original artist's intent. 
S9 too expressed her reservations about making changes because she felt that her changes may not express the original artist's intent, ``\textit{How can we improve their work? I felt that, when drawing, that it requires a lot of effort to know where to fill} [on their work].''

These strong feelings suggested that system design in this area cannot assume that everything is or will be shared, but needs to pay particular attention to ownership, even when an effect of this sentiment is to reject an ethos of reusing and remixing.

\subsubsection{Negotiating Collective Practices}

In addition to the collective rule to not draw over others' drawings, the sister-survivors created other rules for the space.
For example, on the fourth day, when we were discussing the drawing page, S5 said that she liked everything.
To this, S9 proposed a rule, ``\textit{You can't like everything. There has to be something that you don't like.}'' 
We noticed that when a sister-survivor did not follow the rule, it was explicitly mentioned.

Similarly, the sister-survivors negotiated and arrived at mutually accepted elements that they wanted to include in the item-description video:
\begin{quote}
    S6: \textit{I say, ``Namaskar} [Nepali greeting]\textit{, this pote''?}
    
    S7: \textit{Don't say namaskar.}
    
    S6 [asking the group]: \textit{What all should we say?}
    
    S7 [to the group]: \textit{Here it }[the web interface] \textit{says ``talk about your craft.}''
    
    S6: So far, we have ``\textit{This is a pote called Chuche Pote.}''
    
    S8: \textit{Say the name and explain how you made it.}
    
    S6: \textit{So, I say, ``This is a pote called Chuche Pote. It takes 5 hours to make this and ... ''?}
    
    S2: \textit{Does it discolor or not? I don't think it does.}
    
    S6: \textit{This does not discolor. It will last for a long time. And what else do I say? }
    
    S7: \textit{Price? Let's say 300.}
\end{quote}

We find that the sister-survivors followed the template in recording details, as we hear in S3's description:
\begin{quote}
    S3: \textit{This is a bracelet. This is worn on the hand. It takes around 20 minutes to make this. We take a wire and we place Pote on it. We can have in different colors and also in a single color. The price of this is 350. Thank you.}
\end{quote}
30 of the 47 item-description videos described the process of making the handicrafts; 21 of them mentioned a price which ranged from NPR 150 (\textasciitilde USD 1.3) to NPR 1300 (\textasciitilde USD 11.7).

Negotiation of practices around Hamrokala became an avenue for the sister-survivors to leverage their existing social bonds, creating a context for interactions that were not just confirmatory. These incidents and behaviors suggest that conditions allowed room for experimentation about intertwinement of the social and the technical.

\subsection{Crafting}
The current work uncovered a complex story of attachments and also reluctance related to crafting. 
The sister-survivors mentioned diverse aspirations; none of those involved crafting as the primary source of livelihood.
Yet, during the sessions, the sister-survivors' interaction suggested that (1) crafting facilitated the use of the web application, (2) computing supported the growth in their understanding of the crafting practices, eliciting expressions of lack of control and limitations over some aspects of crafting.  

\subsubsection{Computing Introduction Facilitated by Crafting}

The sister-survivors were familiar with buying and selling of the handicrafts. It was a practice that was taught by SO and they had seen SO sell handicrafts to visitors. 
The fact that the web application supported selling was well understood as we heard in S1's item-description video, said, ``\textit{I am putting this here }[on sale]\textit{ in the hope that you all will buy it.}''

Having the web application contextualized around crafting facilitated the understanding and use of the application.  
For example, S1 understood that Hamrokala helped her post items for sale whereas S7 thought it was about ``\textit{sending object} [craft] \textit{information from one place to another}''. These expressions denoted different ways of appreciating the system all of which, while partial, were connected to sharing and selling of crafts.

\subsubsection{Deepening Their Understanding of the Crafting Practices}

While the sister-survivors knew that they had to communicate details of the craft to sell it, they were initially unclear about \emph{what} needed to be communicated. 
The discussion between S2, S6, and S7, as shared above, presenting \textit{Pote} without a greeting remark and by mentioning the name and explaining the process, illuminates a step towards their growth in situating their crafting practices within the wider economic realities.  

Similarly, the computing activities, particularly the creation of item-description videos,  created opportunities for the sister-survivors to reflect on their crafting practices. 
For example, we observed S6 and S7 discuss how long it takes to make a \textit{Pote} leading S6 to exclaim, \textit{My god! It takes 9 hours!} 
A similar discussion ensued because S8 was unfamiliar with how much a Saori scarf, a Japanese-style crocheted scarf, costs:  

\begin{quote}
    S8 [to S9]: \textit{How much would its} [Saori scarf] \textit{price be?} 
    
    S9 [to the table]: \textit{How much would this cost? Around 500-600} [rupees]\textit{?}
    
    S8: \textit{This costs 500-600!}
    
    S7: \textit{Around 1000, I think.}
    
    S8: \textit{Aama!} [exclaiming shock]
\end{quote}

S8, who had worked on Saori scarfs, found the price to be unexpectedly high. Her shock, in fact, can be heard in her eventual item-description video where she says, ``\textit{... the price of this }[scarf] \textit{is very expensive.}'' 

This interaction was one of several that made us aware of questions to be addressed in future interactions with SO about the sister-survivors' time and finances related to crafting.  We were told that about 60\% of the support for SO comes from donations, but we do not know whether income from crafting provides a significant portion of the operating expenses. If it does, the sister-survivors do not seem to be aware of it.

\subsubsection{Limitations Around Crafting} 

Not all the discussions broadened their perspective to situate crafting within the wider economic realm. In some cases, SO's role in managing sales was prominent.
The sister-survivors, like S8, were unaware of the selling price of the crafts they were making. 
While 21 item-description videos quoted a price, they mentioned that they were not aware of the actual selling price and mentioned that they estimated it in the videos.

The sister-survivors also expressed a lack of control over the crafting practice. 
S9 mentioned the power that the trainers hold in deciding the crafting work, ``\textit{With her }[trainer's] \textit{permission and only when she says yes, we start working on it }[a craft].'' 
She added, `` \textit{Do it like this, they } [trainer] \textit{say. They give us the design and we do it looking at the design.}''
On Day 7, they were watching videos of experts making handicrafts that were similar to the ones they make. 
A particular video 
 was well-liked by the entire group. 
They felt that they could make it but were hesitant to try without the trainer's permission.

The centrality of SO and the lack of control over the craft manifested in the sister-survivors audio comments as well. 
None of the 45 comments that the sister-survivors left for each other mentioned a critique or suggested changes on the crafts. 
In contrast, all the comments provided suggested improvements to the description of the item or the video presentation such as S6's comment on S1's craft: ``\textit{The Pote that you have made is nice but you did not show the Pote properly. If only you could bring it up in the front and show it.}'' 
S1 made the suggested change and responded to S6, ``\textit{With your } [comment]\textit{, we realized the shortcoming and we have made changes. It }[your comment] \textit{was helpful.}'' 
Through these suggestions, the sister-survivors seem to express greater power to bring about change in the digital artifact in contrast to the physical ones.

\section{Discussion}
We came into the study with the goals of learning more about the sister-survivors and exploring the possibility of using technology to build upon their pre-existing strengths, namely the social bond between them and their knowledge of crafting. 
We tried to do so by developing and introducing Hamrokala, a web application contextualized around crafting. 
We asked how this particular intervention facilitated, strengthened and extended their strengths, and how it needed to be changed.

The voice annotation in Hamrokala, including the elaborated, colloquial phrasing in a familiar Nepali tone seemed to facilitate initial interaction, even for sister-survivors with low literacy. 
We observed practices of help-seeking and collective problem solving that were similar to those we had observed around crafting earlier \cite{redacted}.
We also observed the development of new communal behaviors in response to the new context.  
Importantly, this encounter was pleasurable or even convivial in Johri and Pal's terms \cite{johri2012capable} and did not appear to leave them disappointed.

In particular, sister-survivors collectively decided together what was to be done with Hamrokala. 
They established and enforced norms and practices such as not being allowed to like everything about the web application or what needs to be said in the item-description video. 
In these ways, they seemed to make the technology their own.    
We also saw emergent behaviors such as reflection about how to describe the artifacts to potential buyers and discussion about the wider socio-economic realities surrounding their crafting.  
The technology was appropriated to build their strengths both directly -- in-so-far as knowledge of technology has power -- and indirectly -- discourse about crafting may be tied to how they see their prospects in the wider socio-economic context.

We do not see technology adoption as an end or even a ``good'' in-and-of-itself, but only in-so-far as it increases existing strengths or assets of the population and, in this case, holds promise of leading to  ``dignified reintegration''. 
Success in this intervention contributes to knowledge about how to conduct asset-based interventions in the design of collaborative socio-technical systems for vulnerable populations.
Our findings elaborate on the detailed knowledge required to tailor the technology and the fit of the technology to the setting to encourage use, appropriation, and the growth of agency in the group.  
Details as small as the use of the word ``comment'' in the interface --- a word that none of the sister-survivors could understand without help --- may determine whether an intervention succeeds or fails.

\subsection{A Case for a Committed Asset-Based Approach}

Assets are not static properties; they are dynamic and are dependent not only on the people but also on the processes and infrastructures available to them. 
Assets can be fraught with tensions as we heard in the sister-survivors' expressions of lack of control over aspects of crafting.
Thus building on assets may require \textit{satisficing} \cite{simon1956rational} across multiple goals and constraints of the situation at different levels \cite{tatar2007design}. 
Moreover, focusing on assets alone may cause us to overlook the power dynamics in the larger system. 
This calls for a commitment to \emph{simultaneously} engage in addressing larger-system issues.

In our analysis, noticing the sister-survivors' ambivalent relationship to crafting in the face of SO's commitment to it and the possible more or less beneficent reasons for that commitment puts us in an ethically fraught position.  
We acknowledge this position but move forward because in the lack of perfect knowledge,  building upon existing strength is more likely to lead to transformational possibility than designing from scratch. 
In taking this stance, we echo Sultana \textit{et al.}'s recommendation to design within a patriarchal society by trying ``to empower women within the structures of their society, instead of trying to destroy those structures.'' \cite[pp. 9]{sultana2018design}. 

We are aware of ``hit-and-run'' academic interventions that ultimately leave gaps upon completion, potentially resulting in increased harm to vulnerable populations and institutions that support them \cite{dearden2012see, dearden2016moving}.
We argue the need for forming and sustaining a long-term relationship and commitment to locating and building upon existing strengths.  
Commitment to work in a setting with vulnerable population calls for restraint on the inclination to promote technology as a solution to intractable problems.  
We advocate modest steps, with particular attention to unexpected negative consequences, an approach that is the opposite of ``move fast and break things'' \cite{taplin2017move}.

\subsection{Using Technology to Create Room for Experimentation}

Much prior work on care emphasizes the relationship between social bonds and the introduction of technology.  It points out that care is a powerful resource but also something that can not be assumed and something that can present difficulties \cite{karusala2017care, mol2010care, murphy2015unsettling}.  
To enact care positively suggests creating room for experimentation  \cite{mol2010care}.
Our data suggest that the sister-survivors experienced room for experimentation in that they adapted the technology in a communal way and created novel use-practices.
We continue to emphasize both that care/communality is an important resource and that it can be built upon with an intervention such as ours, even with such contingent relationships as between the sister-survivors.  

At the same time, our findings show that a communal orientation does not erase individual feelings of ownership and that we must pay attention not only to the communal but also to the boundaries and tensions as individuals experience them. 
The sister-survivors' actions and opinions are multifaceted with room for disagreement, negotiation, and collective and individual ownership as well as displays of care and alignment.

\subsection{Next Steps} 

We shared the findings with SO and four sister-survivors (S2, S5, S8, and S9) who were in the shelter home when we visited SO in August 2019. 
The staff members 
agreed to seek ways to provide greater control over crafting to the sister-survivors.
We plan to move forward in concert.

Given that we can find ways to handle the sister-survivors low text and digital literacy, we believe that computing could be a viable source of strength.  
This opens two related avenues.  
The first is filling in the system and the training in such a way that the sister-survivors can actually sell online.  This would require training in communication, multimedia use, and protecting oneself around computing technology, and establishing infrastructure to connect to a marketplace like Etsy.

The second is to attempt to widen their crafting practices.
We envision communal spaces that afford more chance for the participating women, including the sister-survivors, to innovate, including on the design and production using sewable electronics \cite{buechley2010lilypad}. 
We imagine building on Peppler's work \cite{peppler2014soft, peppler2014short} that specifies patterns for creating soft circuitry.  But these would have to be made culturally appropriate and simpler.

\subsection{Promoting Communal Spaces}
As we contemplate a communal space that would widen the sister-survivors' crafting practices, we believe that, in opposition to the neoliberal ethos and individuality that is fostered in some makerspaces, our approach should involve supporting collectivity and care. 
Collectivity and care have been documented in CHI scholarship (e.g. \cite{fox2015hacking, taylor2016making, toombs2015proper}), but our context would increase these elements both by our design and the sister-survivors' appropriation of the system. 
For example, the making \emph{itself} would be conducted communally. Such a space could augment the sister-survivors' prospects for livelihood both in the context of the protected living situation and perhaps afterward.  It could increase their competitive advantage; however, even if such endeavors were not totally successful in the marketplace, they would give the sister-survivors more experience of their own strength in overcoming obstacles to learning and more confidence in attempting new tasks. 

Three particular design desiderata for the communal space result from the current work that will inform our future, and may be relevant to other people working with vulnerable populations.
These design approaches are not mutually exclusive but rather the provision of each reinforces the other.

\subsubsection{Make Activity Perceptible}

Literature on groupware systems has extensively discussed requirements to maintain group awareness (e.g. \cite{dourish1992awareness, gutwin2002descriptive}).
 A lot of recent systems in affluent circumstances have offered facilities like large displays (e.g. \cite{ fraser2017webuild, hertzum2015visible, lafreniere2016crowdsourced}) that showcase individual accomplishments and indeed invite competition.  
 
We agree with the need for awareness, but a different kind of awareness. 
Sharing needs to be supported to foster communality. 
Voice annotation serves a secondary purpose beside handling low-literacy.  
It also publicizes what someone else is doing or trying to do, making their activity known to others. 
Voice annotation functions as an invitation to other sister-survivors to join in either by the simple display of interest or through the provision of support. 
Further, the sister-survivors integrated both technologically-supported and face-to-face opportunities to show their work to each other into their work practices. 
Facilities that support practices of reporting and sharing are important markers of attention.

\subsubsection{Create Appropriable Steps}

Hamrokala, especially through voice annotation, provided a way for all the sister-survivors to be able to contribute from the start.  
That was important but more important was that someone in the group knew enough to help take the next step.  

Appropriation and designs to support appropriation have been explored in prior research (e.g. \cite{dourish2003appropriation, ludwig2018designing, salovaara2011appropriation}).
Dix \cite{dix2007designing} presents a non-exhaustive list of principles to design for appropriation which includes making the system visible, exposing the intentions behind the system, and encouraging sharing of the appropriated technology. 
We endorse these principles to designing for appropriation but they are primarily individualistic.  

In a communal space, each step in instruction and support must be thought out.  
Not everyone will absorb information in the same time frame \cite{bereiter2005education, stahl2005group}.  
The important elements are the collective knowledge of the group and their ability to put their knowledge together. 
An element is appropriable if some set of people in the group understands enough of its facets and there is enough time to stitch the knowledge together into collectively-meaningful action.  
Such activities enable progress with local tasks and also strengthen social bonds.

\subsubsection{Build in Fun Choices}

We observed the sister-survivors negotiating ownership, practices, and norms within the space. Arguably, part of what enabled this was that the activities were in some sense fun. Communality both enabled this to be fun and was reinforced by it. Fun in this sense is deeply tied to Mol \textit{et al.}'s \cite{mol2010care} emphasis on room for experimentation.  

As we think about the pragmatics of sustaining communal crafting spaces, economic realities will play a significant role. However, we believe that there can be room for playfulness \cite{ferreira2015play, sey2014all, smyth2010there} and more choices. 

In our context, one avenue to fun might be through encouraging sister-survivors to sketch and sew motifs on knitted products and add smaller patterns on \textit{Pote} and bags. Another way is to widen the sister-survivors' modes of expression in using the interface such as by creating and using avatars overlaid on item-description videos. This would show their relationship to the crafts without displaying their real identities. We imagine the creation of such avatars as a communal enterprise.

\subsection{Limitations}
We make small moves because the repercussions of moving fast, in this context, are potentially harmful to an already-vulnerable group. 
Furthermore, with each small move, we deepen our understanding of the possibilities and concerns surrounding the sister-survivors' reintegration journey.  
Hamrokala and the related workshop do not attempt to establish an entire pathway out of poverty and dependence for sex-trafficking survivors in Nepal.  All this study does is begin to establish that some movement may be possible.

Our success in the current study needs to be replicated and extended. A ten-day intervention, even building on another ethnographic study conducted over a more sustained time-frame, does not allow researchers to fully comprehend the complexities and uncertainties in the lives of the sister-survivors or the NGOs that support them.  It is unclear, for example, whether the group reported in earlier work, which had lower literacy levels, would have met with similar success and displayed similar agency as this group.

Novelty effects can be significant \cite{consolvo2008activity}. For now, we build on novelty. We believe that if the sister-survivors had more exposure to computers, the novelty effects would be less, but their agency would be greater and their fear would be reduced.

\section{Conclusion}
\label{con}

We see the NGO's rehabilitation program and the subsequent reintegration process as an on-going journey towards long-term reintegration for these survivors and others in the future. 
We situate our work as an attempt to support the survivors in their own journey which undoubtedly will be varied.
We took a small step in this direction by presenting a voice-annotated web application to a group of survivors. 

The existing strengths --- their crafting skills and their social ties with one another --- were utilized in making the technology and the activities approachable. 
In turn, they were reinforced through the survivors' use of the technology and participation in the activities. 
Knowledge of the computing technology could be another potential resource.

In focusing on communality around the introduction of technology, we hope to further CHI's interest in community building and engagement. 
In particular, we argue for an asset-based, communality-centered and committed approach. 
In doing so, we stand in solidarity with the CHI community to reiterate the critical need for focus on the larger social context within which our designed technologies operate. 

\begin{acks}
We wish to thank the sister-survivors and SO staff members for their time and support, and Muna Khatiwada for lending her voice to Hamrokala.
\end{acks}

%
%
%
%
%
\balance{}

\balance{}


\bibliographystyle{SIGCHI-Reference-Format}
\bibliography{proceedings}

\end{document}